\documentclass[aps,pre,twocolumn,groupedaddress]{revtex4-1}

\usepackage{amsmath}
\usepackage{txfonts}
\usepackage{braket}
\usepackage{graphicx}

\begin{document}


\title{Difference between quantum annealing \\by imaginary-time and real-time Schr\"{o}dinger equation of Grover's search}
\author{Shuntaro Okada$^{1,2}$}
\author{Masayuki Ohzeki$^1$}
\author{Kazuyuki Tanaka$^1$}
\affiliation{$^1$Graduate School of Information Sciences, Tohoku University, Sendai 980-8579, Japan}
\affiliation{$^2$Advanced Research and Innovation Div. 3, DENSO CORPORATION, Kariya, 448-8661, Japan}
\date{\today}

\begin{abstract}
We confirmed the annealing time of Grover's search which is required to obtain desired success probability for quantum annealing by the imaginary-time and the real-time Schr\"{o}dinger equation with two kinds of schedulings;
one linearly decreases the quantum fluctuation and the other tunes the evolution rate of the Hamiltonian based on the adiabatic condition.
With linear scheduling, the required annealing time for quantum annealing by the imaginary-time Schr\"{o}dinger equation is of order $\log N$,
which is very different from $O(N)$ required for the quantum annealing by the real-time Schr\"{o}dinger equation.
With the scheduling based on the adiabatic condition, the required annealing time is of order $\sqrt{N}$,
which is identical to the annealing time for quantum annealing by the real-time Schr\"{o}dinger equation.
Although the scheduling based on the adiabatic condition is optimal for the quantum annealing by the real-time Schr\"{o}dinger equation,
it is inefficient for the quantum annealing by the imaginary-time Schr\"{o}dinger equation.
This result implies that the optimal scheduling for the quantum annealing by the imaginary-time and the real-time Schr\"{o}dinger equation is very different,
and the efficient scheduling considered with the quantum Monte Carlo methods, which is based on imaginary-time Schr\"{o}dinger equation, is not necessarily effective to improve the performance of quantum annealing by the real-time Schr\"{o}dinger equation.
We discuss the efficient scheduling for quantum annealing by the imaginary-time Schr\"{o}dinger equation with respect to the exponential decay of excited states.
\end{abstract}
\maketitle

\section{Introduction}
Combinatorial optimization problems, which are used to minimize cost functions with descrete variables, have significant real-world applications.
Combinatorial optimization problems can generally be mapped to the search of the ground state of a classical Ising Hamiltonian \cite{Ising_mapping}.
Quantum annealing (QA)\cite{QA_original}, which is strongly related to adiabatic quantum computation \cite{AQC_original}, was proposed
as a method for searching the ground state of a Hamiltonian with a complicated energy landscape. This is achieved by utilizing quantum fluctuation to efficiently escape local minima.
Quantum annealing is contrasted with simulated annealing (SA), \cite{SA_original} which employs thermal fluctuations.
Numerous studies have investigated whether QA significantly outperforms SA in terms of the computational time required to obtain a high-accuracy solution.
Most of the studies have shown that QA is superior to SA \cite{QA_sup1, QA_sup2},
while a few have suggested that it is inferior \cite{QA_inf}.
Recently, commercial QA machines based on superconducting flux qubits \cite{D-wave} have been developed by D-Wave Systems Inc.
Experimental researches using the QA machines have been performed, comparing performance of QA and SA \cite{D-wave_SA1, D-wave_SA2}. Other researches have demonstrated the applicability of the QA machines to practical problems \cite{application1, application2}.
In addition, the further improvements of QA machines are discussed deeply.
In particular, one of the promising direction of the improvements is to implement the XX interaction and introduce the non-stoquastic Hamiltonian.
Nontrivial quantum fluctuation, the XX interaction and the other, hamper efficient classical computation due to sign problem except for several examples \cite{OhzekiQMC}.
However the nontrivial quantum fluctuation shows the exponential acceleration for several specific problems \cite{Seki1,Seki2}.
The implementation of the nontrivial quantum fluctuations might be essential to show the quantum supremacy and demonstrate the power of quantum computation. 

On the other hand, quantum annealing using the imaginary-time Schr\"{o}dinger equation (QA-IT) have also been studied, because several computational methods for efficiently investigating quantum dynamical systems are based on imaginary-time Schr\"{o}dinger dynamics.
Stella \textit{et al.} numerically confirmed the asymptotic behavior $1/\tau^2$ of QA-IT, which is identical to QA in the real-time Schr\"{o}dinger equation (QA-RT) for discrete two-level system problems.
Stella validated the practical implementation of QA-RT based on an imaginary-time quantum Monte-Carlo method in the region
where the Landau-Zener transition does not occur both in QA-RT and QA-IT.
In addition, Stella also confirmed that QA-IT provides quantitatively better results than QA-RT during the short time in which the Landau-Zener transition is dominant,
and indicated that a more accurate ground state could be obtained with QA-IT because of the dynamics
which filter the instantaneous ground state efficiently \cite{Simple_case}.
In Ref.\cite{Math_foundation}, the asymptotic behavior $1/\tau^2$ of QA-IT is demonstrated through the imaginary-time version of the adiabatic theorem.
It was also confirmed that QA-IT outperformed QA-RT when applied to the random Ising chain \cite{random_Ising_chain} and $p$-spin infinite-range ferromagnetic Ising model \cite{p_spin_model}.
In Ref.\cite{p_spin_model}, the asymptotic behavior $1/\tau^2$ of QA-IT in the $p$-spin infinite-range ferromagnetic Ising model was also confirmed.

However, to our knowledge, no research has analytically verified the annealing time required to obtain desired success probability for QA-IT.
While the adiabatic conditions for QA-IT are formulated in Ref.\cite{adiabatic_condition},
the annealing time estimated by the adiabatic conditions is very different from that in the numerical results,
because the exponential decay of excited states is not taken into account.
As indicated in Ref.\cite{Simple_case}, if we suppose that the exponential decay of excited states plays an important role in QA-IT,
the adiabatic condition in which the Landau-Zener transition is avoided becomes merely a sufficient condition, rather than a necessary one.

In the present study, we obtain the analytical expression of the annealing time of Grover's search \cite{Grover_original} in QA-IT,
with the fluctuation being linearly decreased.
The obtained annealing time is of order $\log N$,
and is much smaller than that with the fluctuation which is tuned on the basis of the adiabatic condition.
It is well-known that the required annealing time for Grover's search in QA-RT is of order $N$ with the fluctuation being linearly decreased,
and $\sqrt{N}$ with the fluctuation which is optimally tuned on the basis of energy gap \cite{Grover_schedule}.
Our result implies that the efficient scheduling of the quantum fluctuation in QA-IT is very different from that in QA-RT,
and the scheduling tested by quantum Monte Carlo schemes, which is the practical implementations of QA-RT based on imaginary-time Schr\"{o}dinger equation,
may not be effective to improve the performance of QA-RT.
We deal with these differences by considering the exponential decay of excited states in QA-IT.

This paper is organized as follow.
In Sec. 2, we introduce our problem and obtain the analytical expression for the annealing time of Grover's search which is required to obtain desired success probability
in QA-IT with linearly decreasing fluctuations.
In Sec3. we compare the analytical and numerical results.
Sec. 4 presents our discussion and conclusions.

\section{Analytical Results}
In this section, we describe our analytical derivation of the required annealing time of Grover's search in QA-IT with linearly decreasing fluctuations.

The generic form of the time-dependent Hamiltonian for QA is
\begin{equation}
\hat{H}\bigl( s(t) \bigr) = s(t) \hat{H}_0 + \left[ 1 - s(t) \right] \hat{H}_{\rm{q}} ,  \label{eq:QA_Hamiltonian}
\end{equation}
where $\hat{H}_0$ is the classical Hamiltonian, which represents the cost function to be minimized,
and $\hat{H}_{\rm{q}}$ is the quantum fluctuation whose ground state is trivial.
At the beginning of QA, $s(0)$ is set to $0$ and the system is in a trivial ground state determined by the quantum fluctuation.
If we increase $s(t)$ to $1$ sufficiently slowly, the system will remain close to the instantaneous ground state of the time-dependent Hamiltonian.
Thus, we will ultimately obtain the ground state of the classical Hamiltonian $\hat{H}_0$, which represents the optimal solution.
In QA-RT, the system evolves according to the real-time Schr\"{o}dinger equation
\begin{equation}
i \frac{d}{dt} \ket{\psi(t)} = \hat{H}\bigl( s(t) \bigr) \ket{\psi(t)} ,
\end{equation}
while in QA-IT, the system evolves according to the imaginary-time Schr\"{o}dinger equation
\begin{equation}
- \frac{d}{dt} \ket{\psi(t)} = \hat{H}\bigl( s(t) \bigr) \ket{\psi(t)}  \label{eq:IT_shcrodinger},
\end{equation}
where $\ket{\psi(t)}$ is the state vector of the system, and we let $\hbar=1$ for simplicity.
This study investigates QA-IT with linearly decreasing fluctuation, as shown below
\begin{equation}
s(t) = \frac{t}{\tau}  \label{eq:schedule},
\end{equation}
where $\tau$ is the annealing time.

\subsection{Eigenvalues and Eigenvectors of Grover's Search}
The classical Hamiltonian $\hat{H}_0$ and quantum fluctuation $\hat{H}_{\rm{q}}$ of Grover's search \cite{Grover_schedule} are
\begin{gather}
\hat{H}_0 = \hat{I}_N - \ket{0} \bra{0} ,  \label{eq:Grover_H0}  \\
\hat{H}_{\rm{q}} = \hat{I}_N - \ket{\Psi_0} \bra{\Psi_0},  \label{eq:Grover_Hq}
\end{gather}
where $\hat{I}_N$ is the identity operator whose dimension is $N$.
The state vector $\ket{\Psi_0}$ is defined as
\begin{equation}
\ket{\Psi_0} = \frac{1}{\sqrt{N}} \sum_{i=0}^{N-1} \ket{i}.
\end{equation}
The state vector $\ket{0}$ represents the optimal solution, and $N$ is the number of items in the database.
By selecting $\ket{0}$ and
\begin{gather}
\ket{\Psi} = \frac{1}{\sqrt{N-1}} \sum_{i=1}^{N-1} \ket{i},  \\
\ket{\varphi_\alpha} = \sqrt{ \frac{1}{\alpha ( \alpha + 1 )}} \left[ \sum_{i=1}^{\alpha} \ket{i} - \alpha \ket{\alpha+1} \right],
\end{gather}
as the bases of the system, the Hamiltonian $\hat{H}\bigl( s(t) \bigr)$ can be written as follows.
\begin{equation}
	\hat{{H}}\bigl( s(t) \bigr) = \left[
	\renewcommand{\arraystretch}{1.5}
	\begin{array}{@{\,}cc|ccc@{\,}}
	\displaystyle \bigl( 1 - s(t) \bigr) \left( 1 - \frac{1}{N} \right) & \displaystyle - \bigl( 1 - s(t) \bigr) \frac{\sqrt{N-1}}{N} &    \\[15pt]
	\displaystyle - \bigl( 1 - s(t) \bigr) \frac{\sqrt{N-1}}{N} & \displaystyle \bigl( 1 - s(t) \bigr) \frac{1}{N} + \Gamma &    \\[5pt]  \hline
	   &   & \hat{I}_{N-2}
	\end{array}
	\renewcommand{\arraystretch}{1}
	\right].
\end{equation}
If the initial state is chosen to be the trivial ground state of $\hat{H}_{\rm{q}}$,
then the dynamics described by $\hat{H}\bigl( s(t) \bigr)$ are restricted to the state space spanned by $\ket{0}$ and $\ket{\Psi}$.
The eigenvectors and eigenvalues that contribute to the instantaneous state in QA are given by
\begin{gather}
\ket{0 \bigl( s(t) \bigl)} =   P \bigl( s(t) \bigl) \ket{0} + Q \bigl( s(t) \bigl) \ket{\Psi} ,  \label{eq:eigen_ground}  \\
\ket{1 \bigl( s(t) \bigl)} = -Q \bigl( s(t) \bigl) \ket{0} + P \bigl( s(t) \bigl) \ket{\Psi} ,  \label{eq:eigen_excited}  \\
\intertext{where}
P \bigl( s(t) \bigl) = \sqrt{\frac{1}{2} + \left[ \frac{1}{2} - \left( 1- \frac{1}{N} \right) \bigl( 1 - s(t) \bigr) \right] \frac{1}{\Delta \varepsilon_{10} \bigl( s(t) \bigl)}},  \\
Q \bigl( s(t) \bigl) = \sqrt{\frac{1}{2} - \left[ \frac{1}{2} - \left( 1- \frac{1}{N} \right) \bigl( 1 - s(t) \bigr) \right] \frac{1}{\Delta \varepsilon_{10} \bigl( s(t) \bigl)}},  \\
\intertext{and}
\varepsilon_0 \bigl( s(t) \bigl) = \frac{1}{2} \left[ 1 - \Delta \varepsilon_{10} \bigl( s(t) \bigl) \right],  \\
\varepsilon_1 \bigl( s(t) \bigl) = \frac{1}{2} \left[ 1 + \Delta \varepsilon_{10} \bigl( s(t) \bigl) \right].
\end{gather}
Here $\Delta \varepsilon_{10} \bigl( s(t) \bigr)$ is the energy gap given by
\begin{equation}
\Delta \varepsilon_{10} \bigl( s(t) \bigl) = \sqrt{ 1 - 4 \left( 1 - \frac{1}{N} \right) s(t) \bigl( 1 - s(t) \bigr)}.
\end{equation}
The matrix element of $d \hat{H} \bigl( s(t) \bigr) / dt$, which we will use later, is written as
\begin{equation}
\braket{1 \bigl( s(t) \bigr)|\frac{d \hat{H} \bigl( s(t) \bigr)}{d t}|0 \bigl( s(t) \bigr)} = \frac{1}{\tau}\frac{\sqrt{N-1}}{N} \frac{1}{\Delta \varepsilon_{10}(\Gamma)}  \label{eq:matrix_element},
\end{equation}
because the schedule for $s(t)$ is given by Eq. (\ref{eq:schedule}).

\subsection{Coefficient of Excited State}
Let us assess the upper bounds of the coefficient for the excited state.

Because the schedule for $s(t)$ is given by Eq. (\ref{eq:schedule}),
the imaginary-time Schr\"{o}dinger equation (\ref{eq:IT_shcrodinger}) is rewritten in terms of $s$ as
\begin{equation}
- \frac{1}{\tau} \frac{d}{ds} \ket{\psi(s)} = {\hat{H}}(s) \ket{\psi(s)}.  \label{eq:IT_shcrodinger_Gamma}
\end{equation}
Following Ref.\cite{adiabatic_condition}, we expand the state vector in terms of the set of instantaneous eigenstates as
\begin{gather}
\ket{\psi(s)} = \sum_{k=0}^{1} C_k (s) \ket{k(s)} = \sum_{k=0}^{1} \tilde{C}_{k}(s) e^{-\tau \phi_k (s)} \ket{k (s)},  \label{eq:expand_psi}  \\
\intertext{where}
\phi_k (s) = \int_{0}^{s} \varepsilon_{k} (s') ds'  \label{eq:phi_integral}.  \\
\intertext{Here we define}
\hat{H}(s) \ket{k(s)} = \varepsilon_{k}(s) \ket{k (s)}.
\end{gather}
By substituting Eq. (\ref{eq:expand_psi}) into the imaginary-time Schr\"{o}dinger equation (\ref{eq:IT_shcrodinger_Gamma}), and combining that equation with Eq. (\ref{eq:matrix_element}),
we obtain differential equations for the coefficients as
\begin{gather}
\frac{d\tilde{C}_{1}(s)}{ds} = + \frac{\sqrt{N-1}}{N} \frac{\tilde{C}_{0}(s)}{\Delta \varepsilon_{10}(s)^2} 
	e^{+ \tau \Delta \phi_{10}(s)}  \\
\frac{d\tilde{C}_{0}(s)}{ds} = - \frac{\sqrt{N-1}}{N} \frac{\tilde{C}_{1}(s)}{\Delta \varepsilon_{10}(s)^2}
	e^{- \tau \Delta \phi_{10}(s)},  \\
\intertext{where}
\Delta \phi_{10}(s) = \int_{0}^{s} \Delta \varepsilon_{10} (s') ds'.  \label{eq:gap_integral}
\end{gather}
Integrating one of the differential equations yields
\begin{equation}
\tilde{C}_{1} (s) = \tilde{C}_{1} (0) + \frac{\sqrt{N-1}}{N} \int_{0}^{s} ds' \frac{\tilde{C}_{0}(s')}{\Delta \varepsilon_{10}(s')^2}
	e^{\tau \Delta \phi_{10}(s')}  \label{eq:tilde_C}.
\end{equation}
If the initial condition is chosen to be $\tilde{C}_{0}(0)=C_{0}(0)=1$ and $\tilde{C}_{1}(0)=C_{1}(0)=0$,
the upper bound of $\tilde{C}_{0}(s)$ is $1$ because $d\tilde{C}_{0}(s) / ds \leq 0$.
Substituting the initial conditions and upper coefficient bound into Eq. (\ref{eq:tilde_C}) and multiplying the resulting expression by $ e^{-\tau \phi_1 (s)}$ yields
\begin{align}
&C_{1}(s) \leq e^{-\tau \phi_{0}(s)} D_{1}(s)  \\
&D_{1}(s) \equiv \frac{\sqrt{N-1}}{N} e^{-\tau \Delta \phi_{10}(s)} \int_{0}^{s} \frac{ds'}{\Delta \varepsilon_{10}(s')^2}
	e^{\tau \Delta \phi_{10}(s')}.  \label{eq:D_definition}
\end{align}
The asymptotic behavior of long-term annealing has been already obtained in Refs.\cite{Math_foundation} and \cite{adiabatic_condition}.
Here, we calculate the upper bound of $D_{1}(s)$ for short-term annealing.
In deriving asymptotic behavior, integration by parts yields multiplication by $1/\tau$,
because $e^{\tau \Delta \phi_{10}(s')}$ in the integrand of Eq. (\ref{eq:D_definition}) is integrated.
Consequently, the resulting expression is accurate only if the annealing time is sufficiently long.
We carry out the integration by parts by differentiating $e^{\tau \Delta \phi_{10}(s')}$,
which yields the multiplication by $\tau$. The resulting expression is accurate only if the annealing time is sufficiently short.
To simplify the following calculations, we first carry out integration by parts, using the same method which is used to derive the asymptotic behavior.
\begin{multline}
D_{1}(s) = \frac{1}{\tau} \frac{\sqrt{N-1}}{N} e^{-\tau \Delta \phi_{10} (s)}
	\left\{ \left[ \frac{1}{\Delta \varepsilon_{10} (s')^3} e^{\tau \Delta \phi_{10} (s')} \right]_{s'=0}^{s'=s} \right.  \\
\left. + 12 \left( 1- \frac{1}{N} \right) \int_{0}^{s} \frac{ds'}{\Delta \varepsilon_{10}(s')^5} \left( s' - \frac{1}{2} \right) e^{\tau \Delta \phi_{10} \left( s' \right)} \right\}.
\label{eq:integral_by_part1}
\end{multline}
When $N \gg 1$, the integration of Eq. (\ref{eq:gap_integral}) and approximation of the resulting expression yields
\begin{equation}
\Delta \phi_{10}(s) \simeq \frac{1}{2} \left( s - \frac{1}{2} \right) \Delta \varepsilon_{10} ( s ) + \frac{1}{4}.  \label{eq:phi_approximation}
\end{equation}
By substituting Eq. (\ref{eq:phi_approximation}) into Eq. (\ref{eq:integral_by_part1}), $D_{1}(s)$ at the end of QA-IT is given by
\begin{multline}
D_{1}(s=1) \simeq \frac{1}{\tau} \frac{\sqrt{N-1}}{N} \left( 1 - e^{-\frac{\tau}{2}} \right)  \\
	+ \frac{12N}{\tau} \left( \frac{\sqrt{N-1}}{N} \right)^3 e^{-\frac{\tau}{4}} I_{1}(1,0),  \label{eq:DtoI}
\end{multline}
and
\begin{equation}
I_{1}(s_b,s_a) \equiv \int_{s_a}^{s_b} \frac{ds}{\Delta \varepsilon_{10}(s)^5} \left( s - \frac{1}{2} \right) 
	e^{\frac{\tau}{2} \left( s - \frac{1}{2} \right) \Delta \varepsilon_{10}(s)}.
\end{equation}
In addition, because we are interested in the required annealing time at $N \gg 1$,
we use the approximation shown below:
\begin{equation}
\frac{\tau}{2} \left( s - \frac{1}{2} \right) \Delta \varepsilon_{10}(s) \simeq \tau \left( s - \frac{1}{2} \right) \left| s - \frac{1}{2} \right|.
\end{equation}
By using these approximations, we calculate the upper bound of $I_{1}(1,0)$.

First, we calculate the upper bound of $I_{1}(1/2,0)$.
Transforming the integration variable to $x=4(N-1)(s -1/2)^2+1$ yields
\begin{gather}
I_{1} \left( \frac{1}{2}, 0 \right) \simeq - \frac{1}{8} \frac{N^{\frac{5}{2}}}{N-1} e^{\frac{1}{4} \frac{\tau}{N-1}} J_{1} \left( N, 1 \right),  \\
J_{1} \left( N, 1 \right) \equiv \int_{1}^{N} \frac{dx}{x^{\frac{5}{2}}} e^{-\frac{x}{4} \cdot \frac{\tau}{N-1}}.
\end{gather}
It is straightforward to iteratively integrate $1/x^{5/2}$. When we integrate by parts three times, we obtain
\begin{multline}
J_{1} \left( N, 1 \right) = \frac{2}{3} e^{- \frac{1}{4} \frac{\tau}{N-1}} \left[ 1 - \frac{1}{2} \left( \frac{\tau}{N-1} \right) - \frac{1}{4} \left( \frac{\tau}{N-1} \right)^2 \right]  \\
	- \frac{2}{3} \left( \frac{1}{N} \right)^{\frac{3}{2}} e^{-\frac{1}{4} \frac{N\tau}{N-1}} \left[ 1 - \frac{1}{2} \left( \frac{N\tau}{N-1} \right) - \frac{1}{4} \left( \frac{N\tau}{N-1} \right)^2 \right]  \\
	+ \frac{1}{24} \left( \frac{\tau}{N-1} \right)^3 \int_{1}^{N} \sqrt{x} e^{-\frac{x}{4} \frac{\tau}{N-1}} dx  \label{eq:J_by_part}.
\end{multline}
The following inequality
\begin{equation}
\frac{1}{1+\sqrt{N}}x + \frac{\sqrt{N}}{1+\sqrt{N}} \leq \sqrt{x},
\end{equation}
which can be validated for $1<x<N$, is substituted into the third term in Eq. (\ref{eq:J_by_part}). This gives the lower bound of $J_{1}(N,1)$ and the resulting upper bound of $I_{1}(1/2,0)$.
\begin{multline}
I_{1} \left( \frac{1}{2}, 0 \right) \leq -\frac{1}{12} \frac{N^{\frac{5}{2}}}{N-1} \left[ 1 - \frac{1}{2} \frac{\sqrt{N}-1}{\sqrt{N}+1} \left( \frac{\tau}{N-1} \right) \right]  \\
	+ \frac{1}{12} \frac{N}{N-1} e^{-\frac{\tau}{4}} \left[ 1 + \frac{1}{2} \frac{\sqrt{N}-1}{\sqrt{N}+1} \left( \frac{N\tau}{N-1} \right) \right].  \label{eq:upper_bound1}
\end{multline}
The upper bound of $I_{1}(1,1/2)$ is obtained with the same method used above to derive the upper bound of $I_{1}(1/2,0)$,
and is given by
\begin{multline}
I_{1}\left( 1, \frac{1}{2} \right) \leq -\frac{1}{12} \frac{N}{N-1} e^{\frac{\tau}{4}} \left[ 1 - \frac{1}{2} \frac{\sqrt{N}-1}{\sqrt{N}+1} \left( \frac{N\tau}{N-1} \right) \right]  \\
	+ \frac{1}{12} \frac{N^{\frac{5}{2}}}{N-1} \left[ 1 + \frac{1}{2} \frac{\sqrt{N}-1}{\sqrt{N}+1} \left( \frac{\tau}{N-1} \right) \right]  \label{eq:upper_bound2}.
\end{multline}
From Eqs. (\ref{eq:DtoI}), (\ref{eq:upper_bound1}), and (\ref{eq:upper_bound2}), we can obtain the upper bound of $D_{1}(s=1)$, as shown below.
\begin{align}
D_{1} \left( s=1 \right) &\leq \frac{1}{2\sqrt{N-1}} \frac{\sqrt{N}-1}{\sqrt{N}+1} \left( 1 + 2 \sqrt{N} e^{-\frac{\tau}{4}} + e^{-\frac{\tau}{2}} \right)  \notag \\
&\simeq \frac{1}{2\sqrt{N}} + e^{-\frac{\tau}{4}}.  \label{eq:excited_exponential}
\end{align}
The second term indicates the exponential decay of the excited state coefficient (without size dependence).
In contrast, the coefficient in QA-RT remains at approximately $1$ until the adiabatic condition is satisfied.
This makes the required annealing time of Grover's search for QA-IT much smaller than that for QA-RT, as shown below.

The asymptotic behavior at the limits of adiabatic evolution is given by the first term of Eq. (\ref{eq:integral_by_part1}).
\begin{equation}
D_{1}(s=1) \simeq \frac{1}{\tau} \frac{\sqrt{N-1}}{N} \left( 1 - e^{-\frac{\tau}{2}} \right).  \label{eq:excited_asymptotic}
\end{equation}
The probability of finding the excited state is given by the square of the coefficient of the excited state.
Equation (\ref{eq:excited_asymptotic}) indicates that the asymptotic behavior of QA-IT in Grover's search is $1/\tau^2$,
which is consistent with Refs.\cite{Math_foundation} and \cite{adiabatic_condition}.

\subsection{ Ground State Coefficient}
In addition, we must obtain the coefficient of the ground state, because QA-IT does not conserve the norm of the state vector.
We evaluate the lower bounds of the ground state coefficient using both the upper bound of the excited state coefficient and
the lower bound of the success probability, which is the probability of obtaining the optimal solution.

First, we verify
\begin{equation}
\frac{d}{dt} \sqrt{P_{\rm{opt}}(t)} = \frac{d}{dt} \frac{ \braket{0|\psi(t)} }{\sqrt{\braket{\psi(t)|\psi(t)}}} \geq 0,  \label{eq:destination_inequality}
\end{equation}
where $P_{\rm{opt}}$ is the probability of obtaining the optimal solution, and $\ket{0}$ is the optimal solution.
This inequality means that the success probability of the final state in QA-IT is higher than that of the initial state.
The time-dependence of the norm in QA-IT is given by
\begin{equation}
- \frac{d}{dt} \braket{\psi(t)|\psi(t)} = 2 \braket{\psi(t)|\hat{H} (t)|\psi(t)}  \label{eq:norm_dependence},
\end{equation}
and the substitution of Eq. (\ref{eq:norm_dependence}) into Eq. (\ref{eq:destination_inequality}) yields
\begin{multline}
\frac{d}{dt} \sqrt{P_{\rm{opt}}(t)} =  \\
\frac{ \braket{0|\psi(t)} \braket{\psi(t)|\hat{H}(t)|\psi(t)} - \braket{\psi(t)|\psi(t)} \braket{0|\hat{H}(t)|\psi(t)} }
	{ \braket{\psi(t)|\psi(t)}^{\frac{3}{2}} }.  \label{eq:Popt_differential}
\end{multline}
By using Eqs. (\ref{eq:eigen_ground}) and (\ref{eq:eigen_excited}), $\ket{0}$ can be expanded in terms of the set of instantaneous eigenstates as
\begin{gather}
\ket{0} = P(t) \ket{0(t)} - Q(t) \ket{1(t)}  \label{eq:opt_expansion},  \\
0 \leq P(t), Q(t) \leq 1.
\end{gather}
In addition, the state vector can be expanded as
\begin{gather}
\ket{\psi(t)} = L(t) \left[ \alpha(t) \ket{0(t)} + \sqrt{ 1 - \alpha(t)^2 } \ket{1(t)} \right],  \label{eq:psi_expansion}  \\
0 \leq \alpha(t) \leq 1,
\end{gather}
where $L(t)$ is the norm of the state vector.
Substituting Eqs. (\ref{eq:opt_expansion}) and (\ref{eq:psi_expansion}) into Eq. (\ref{eq:Popt_differential}) yields
\begin{multline}
\frac{d}{dt} \sqrt{P_{\rm{opt}}(t)} =  \\
\alpha(t) \sqrt{1-\alpha(t)^2} \left[ \sqrt{1-\alpha(t)^2} P(t) + \alpha(t) Q(t) \right] \Delta \varepsilon_{10}(t) \geq 0,
\end{multline}
and the success probability of the final state satisfies
\begin{equation}
P_{\rm{opt}}(\tau) \ge \frac{1}{N},  \label{eq:Popt_lower}
\end{equation}
where $1/N$ is the success probability of the initial state.

When $\tau \to 0$, Eq. (\ref{eq:excited_exponential}) becomes
\begin{equation}
D_{1}(s=1) \leq 1.  \label{eq:excited_upper}
\end{equation}
We can obtain the lower bound of the ground state coefficient using Eqs. (\ref{eq:Popt_lower}) and (\ref{eq:excited_upper}).
\begin{gather}
C_{0}(s) \equiv e^{-\tau \phi_{0}(s)} D_{0}(s),  \\
D_{0}(s=1) \geq \frac{1}{\sqrt{N}}.  \label{eq:ground_lower}
\end{gather}

\subsection{Required Annealing Time}
We derive the required annealing time of Grover's search.
From Eqs. (\ref{eq:excited_exponential}) and (\ref{eq:ground_lower}), the excited-state-coefficient to ground-state-coefficient ratio can be written as
\begin{equation}
\frac{C_{1}(s=1)}{C_{0}(s=1)} = \frac{D_{1}(s=1)}{D_{0}(s=1)} \leq \frac{1}{2} + \sqrt{N} e^{-\frac{\tau}{4}}.
\end{equation}
The dominant term in $N \gg 1$ is the second term. The annealing time, which is then required to satisfy
\begin{equation}
\sqrt{N} e^{-\frac{\tau}{4}} = \delta,
\end{equation}
is given by
\begin{equation}
\tau = 2 \log \left( \frac{N}{\delta^2} \right).  \label{eq:tau_analytical}
\end{equation}
This result indicates that the annealing time of Grover's search which is required to obtain desired suscess probability is of order $\log N$ in QA-IT with linearly decreasing fluctuation.

\section{Numerical Result}
\begin{figure}
\centering
\includegraphics[width=7cm]{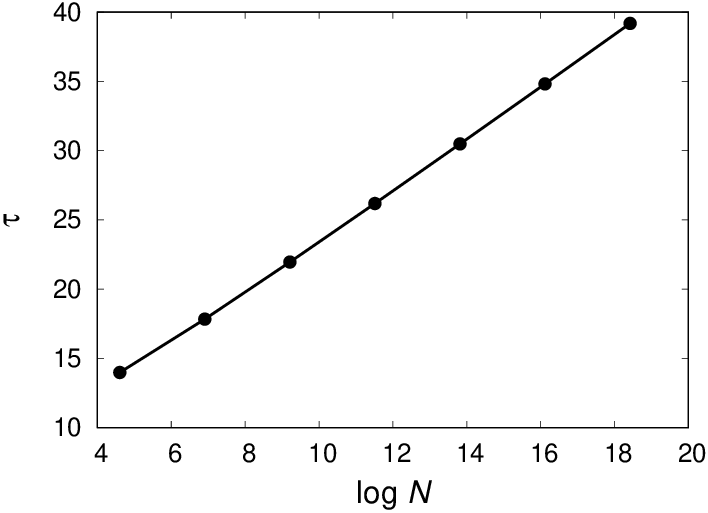}
\caption{$\tau$ for the linear scheduling}
\label{fig:tau_numerical}
\end{figure}

To verify our analytical results, we simulated QA-IT by solving the imaginary-time Schr\"odinger equation using the Runge-Kutta method.
The Hamiltonian is expressed by Eqs. (\ref{eq:QA_Hamiltonian}), (\ref{eq:schedule}), (\ref{eq:Grover_H0}), and (\ref{eq:Grover_Hq}).
We then calculated the success probability of Grover's search at the end of annealing.
The numerical result for the annealing time $\tau$ required to achieve $99\%$ success probability is shown in Fig. \ref{fig:tau_numerical}, where the horizontal axis is $\log N$ and the vertical axis is $\tau$.
The numerical result implies that the $\tau$ required to achieve $99\%$ success probability is  proportional to $\log N$, which is consistent with the analytical results.
The linear fitting of the numerical results provides
\begin{equation}
\tau = 1.83 \log N + 5.27.
\end{equation}
The coefficient of $\log N$ almost coincides with the analytical result, which is $2$ in Eq. (\ref{eq:tau_analytical}), larger than that of the numerical result. 
Lower bound of the ground state coefficient is derived in the limit of $\tau \to 0$, and Eq.(\ref{eq:ground_lower}) is not tight.
This makes the required annealing time $\tau$ large.

\begin{figure}
\centering
\includegraphics[width=7cm]{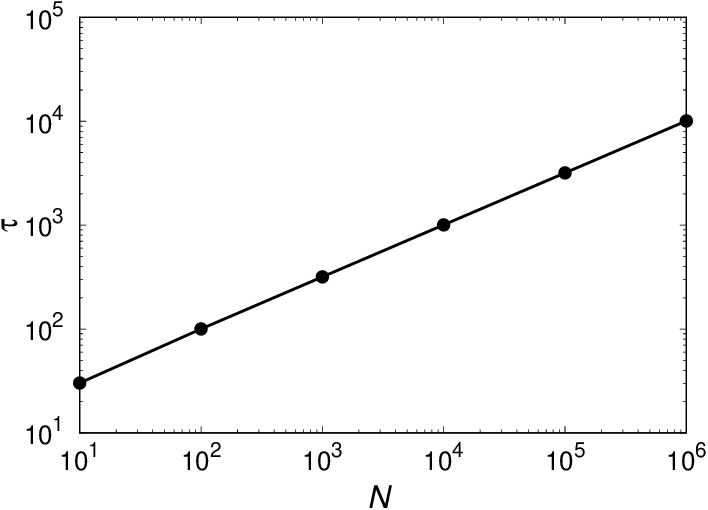}
\caption{$\tau$ for the scheduling based on the adiabatic condition}
\label{fig:tau_numerical_EG}
\end{figure}

We next simulate the annealing time $\tau$ required to achieve $99\%$ success probability for the scheduling as shown below,
\begin{equation}
\frac{ds(t)}{dt} \propto \frac{\Delta \varepsilon (s)^2}{\braket{ 1(s) | \hat{H}_{0} - \hat{H}_{q} | 0(s) } },
\end{equation}
where the evolution rate is adjusted based on adiatabic conditions.
The numerical result is shown in Fig. \ref{fig:tau_numerical_EG}, where the horizontal axis is $N$ and the vertical axis is $\tau$.
The linear fitting of the numerical results provides
\begin{equation}
\log_{10} \tau = 0.50 \log_{10} N + 0.99,
\end{equation}
and the required annealing time is of order $\sqrt{N}$ for the fluctuation based on the adiabatic condition.
Although the scheduling based on the adiabatic condition is optimal for QA-RT,
the linear scheduling is much more efficient for QA-IT.
These results imply that the efficient scheduling is very different between QA-IT and QA-RT,
and the efficient scheduling considered by the quantum Monte Carlo methods
which is based on imaginary-time Schr\"odinger equation is not necessarily effective to improve the performance of QA-RT.

\section{Discussion and Conclusions}
In this paper, we analytically verified that the annealing time which is required to obtain desired success probability is of order $\log N$ with the fluctuation being linearly decreased
and confirmed that the efficient scheduling for QA-IT and QA-RT is very different.
This result indicates the efficient scheduling considered by the quantum Monte Carlo methods which is based on imaginary-time Schr\"odinger equation is not necessarily effective to improve the performance of QA-RT.

With the linear fluctuation, the required annealing time for QA-IT ($O(\log N)$) is much smaller than that for QA-RT ($O(N)$).
This difference is caused by the exponential decay of excited states in QA-IT.
Even if the Landau-Zener transition occurs, we can obtain a high success probability in QA-IT
because the energy gap reopens after the Landau-Zener transition and the exponential decay of excited states play an important role.
With respect to the adiabatic condition, avoiding the Landau-Zener transition is a necessary condition for achieving high success probability in QA-RT.
In contrast, in QA-IT, it is a sufficient rather than necessary condition to avoid the Landau-Zener transition.
As a result, the required annealing time derived from the adiabatic condition \cite{adiabatic_condition} has a different size dependence from the analytical and numerical results in our paper.
It is more important to utilize the exponential decay of excited states than to avoid th Landau-Zener transition for QA-IT.

While, with the fluctuation based on the adiabatic condition, the required annealing time for QA-IT and QA-RT is identical ($O(\sqrt{N})$).
This is because the exponential decay of excited states does not play an important role.
The evolution rate is adjusted to avoid the Landau-Zener transition and pass through the region where the energy gap is large.
As a result, less time is provided for the exponential decay of excited states and it does not contribute to the performance of QA-IT.
With this flucuation, the success probability of QA-IT and QA-RT is identical not only for adiabatically evolved region but also for the short time region where Landau-Zener transition occurs.
Although this scheduling is optimal for QA-RT, it is conjectured that this scheduling is worst for QA-IT in Grover's search.

Finally, we discuss the efficient evolution rate of the Hamiltonian in QA-IT.
In a case where the energy gap monotonically decreases with fluctuation,
the exponential decay of excited states is not expected to play an important role, and it is required to avoid the Landau-Zener transition.
As a result, it is efficient to decrease the evolution rate with respect to the fluctuation (based on the energy gap).
In a case where the energy gap reopens after reaching the minimum value,
it will be more efficient to slowly decrease the fluctuation after reaching the minimum value to utilize the exponential decay of excited states,
rather than slowly decreasing the fluctuation based on the energy gap to avoid the Landau-Zener transition.
In each case, it is efficient to gradually decrease the evolution rate of the Hamiltonian with respect to the fluctuations in QA-IT.
This is very different from QA-RT.
This efficient evolution rate can be also applied to the temperature in SA, because the master equation can be mapped to the imaginary-time Schr\"{o}dinger equation \cite{Master_to_IT}.

\begin{acknowledgments}
The authors are deeply grateful to Manaka Okuyama and Jun Takahashi for their constructive comments and discussions,
which were essential to our research.
Author M. O. is grateful to the financial support from JSPS KAKENHI 15H03699 and 16H04382, the Inamori Foundation, the JST-START and the ImPACT program.
\end{acknowledgments}

\end{document}